\documentclass[11pt]{article}%
\usepackage{moriond,epsfig}
\usepackage{amsmath}
\usepackage{amsfonts}
\usepackage{amssymb}
\usepackage{graphicx}%
\def\be{\begin{equation}}
\def\ee{\end{equation}}
\def\bea{\begin{eqnarray}}
\def\eea{\end{eqnarray}}

\begin{document}
\vspace*{4cm}
\title{Nuclear matter and chiral phase transition at large-$N_{c}$}
\author{Francesco Giacosa}

\address{Institute of theoretical Physics, Johann Wolfgang Goethe University\\
Max-von-Laue Str. 1, D-60438, Frankfurt am Main, Germany}
\maketitle

\abstracts{Two aspects of the QCD phase diagrams are studied in the limit of a large number of colors: at zero
temperature and nonzero density
the (non)existence of nuclear matter, and at zero density and nonzero temperature the chiral phase transition.}

\section{Introduction and Summary}

The limit in which the number of colors $N_{c}$ is sent to infinity
(large-$N_{c}$ limit) represents a systematic approach \cite{witten} to study
properties of QCD. The world for $N_{c}\gg3$ is simpler because planar
diagrams dominate. However, the basic ingredients `survive' in the
large-$N_{c}$ limit: quark-antiquark mesons exist and become weakly
interacting, baryons also exist but are formed of $N_{c}$ quarks. Recently, a
lot of effort has been spent to study the properties of the phase diagram of
QCD when $N_{c}$ is varied \cite{mclerran}.

Along the line of zero temperature and nonzero chemical potential, a natural
question \cite{lucafra} is if nuclear matter binds for $N_{c}>3$. We shall
find that this is not the case: in view of the peculiar nature of the scalar
attraction between nuclei we obtain that nuclear matter ceases to form as soon
as $N_{c}>3$ is considered. Namely, the scaling behavior of the scalar
attraction depends on the nature of the exchanged field with a mass of about
$0.6$ GeV. Present knowledge in low-energy QCD spectroscopy \cite{amslerrev}
shows that this light scalar field is (predominately) not a quark-antiquark
field, the alternative possibilities being tetraquark, pion-pion interpolating
field, molecular state, etc. In all these interpretations the scalar
attraction diminishes in comparison with the vector repulsion, mediated by the
well-known vector meson $\omega$, when $N_{c}$ is increased. As a result,
nuclear matter does not take place \cite{lucafra}: the investigation leading
to this result is achieved though a simple effective model of the Walecka type.

When moving along the finite temperature axis while keeping the density to
zero, it is interesting to study how different chiral effective models behave
at large-$N_{c}.$ It is quite remarkable that two very well-known models, the
quark-based Nambu Jona-Lasinio (NJL) model \cite{njlorig,njl} and the
hadron-based $\sigma$-model \cite{geffen,denis}, deliver different result for
the critical temperature for chiral restoration $T_{c}.$ While in the NJL
model $T_{c}$ scales as $N_{c}^{0}$ and is thus, just as the deconfinement
phase transition, large-$N_{c}$ independent, in the $\sigma$-model one obtains
that $T_{c}\propto\sqrt{N_{c}}$. This mismatch can be solved by including in
the $\sigma$-model one (or more) $T$-dependent parameter(s): a rather simple
modification of the mass term is enough to reobtain the expected scaling
$T_{c}\propto N_{c}^{0}$.

The paper \footnote{Based on the presentation given at `Rencontres de Moriond,
QCD and High Energy Interactions', March 20-27 2011, La Thuile (Italy).} is
organized as follows: in Sec. 2 and Sec. 3 we study nuclear matter and the
chiral phase transition for $N_{c}>3$, respectively. In Sec. 4 we briefly
present our conclusions.

\section{Nuclear matter at large-$N_{c}$}

Nuclear matter at large-$N_{c}$ is studied by means of an effective
Walecka-Lagrangian \cite{walecka}%
\begin{equation}
\mathcal{L}=\bar{\psi}[\gamma^{\mu}(i\partial_{\mu}-g_{\omega}\omega_{\mu
})-(m_{N}-g_{S}S)]\psi+\frac{1}{2}\partial^{\mu}S\partial_{\mu}S-\frac{1}%
{2}m_{S}^{2}S^{2}+\frac{m_{\omega}^{2}}{2}\omega_{\mu}\omega^{\mu}+...
\end{equation}
where $S$ represents a scalar field with a mass of about $0.6$ GeV and
$\omega$ the isoscalar vector meson. The large-$N_{c}$ scaling properties of
the latter are well known: $m_{\omega}\propto N_{c}^{0}$, $g_{\omega}%
\propto\sqrt{N_{c}}$.$\ $We now examine the possibilities \cite{lucafra} for
the scalar state $S$:

$\bullet$ $S$ as quark-antiquark field: $m_{S}\propto N_{c}^{0}$ and
$g_{S}\propto\sqrt{N_{c}}.$ This is the only case in which nuclear matter
exists in the large-$N_{c}$ limit. The binding energy increases with $N_{c}.$
However, this scenario is --as previously anticipated-- unfavored
\cite{amslerrev}.

$\bullet$ $S$ as tetraquark field \cite{jaffeorig}: $m_{S}\propto N_{c}$ and
$g_{S}\propto N_{c}^{0}.$ Nuclear matter does not bind for $N_{c}>3.$ On the
contrary, for $N_{c}=2$ an increased binding is found. This scenario
represents a viable possibility in agreement with phenomenology. It might also
play an important role at nonzero temperature and density \cite{heinz}.

$\bullet$ $S$ as an effective two-pion-exchange effect \cite{weisenucl}:
$m_{S}\sim2m_{\pi}\propto N_{c}^{0},$ $g_{S}\propto\sqrt{N_{c}}.$ Although the
scaling laws are the same as in the quark-antiquark case, no binding is
obtained in view of numerical details.

$\bullet$ $S$ as a low-mass scalar glueball \cite{varieglue}: $m_{S}\propto
N_{c}^{0}$ and $g_{S}\propto N_{c}^{0}.$ No binding for $N_{c}>3$ is obtained.
Note, this scenario is unfavored by present lattice data which place the
glueball at about $1.6$ GeV \cite{lattglue}.

The result that no nuclear matter exists for large-$N_{c}$ is stable and does
not depend on numerical details. In the framework of the so-called strong
anthropic principle it is then natural that we live in a world in which
$N_{c}$ is not large.

\section{Chiral phase transition at large-$N_{c}$}

The $\sigma$-model has been widely used to study the thermodynamics of QCD
\cite{rischkerev}. In one of its simplest forms it reads (as function of
$N_{c}$):%
\begin{equation}
\mathcal{L}_{\sigma}(N_{c})=\frac{1}{2}(\partial_{\mu}\Phi)^{2}+\frac{1}{2}%
\mu^{2}\Phi^{2}-\frac{\lambda}{4}\frac{3}{N_{c}}\Phi^{4}\text{ ,} \label{ls}%
\end{equation}
where $\Phi^{t}=(\sigma,\vec{\pi})$ describes the scalar field $\sigma$ and
the pseudoscalar pion triplet $\vec{\pi}$. The quark-antiquark field $\sigma$
represents the chiral partner of the pion: as mentioned in the previous
section, it does not correspond to the resonance $f_{0}(600)$ with a mass of
about $0.6$ GeV but to the resonance $f_{0}(1370)$ with a mass of about $1.3$
GeV \cite{amslerrev,denis}. The scaling law $\lambda\rightarrow3\lambda/N_{c}$
takes into account that the meson-meson scattering amplitude scales as
$N_{c}^{-1}.$ On the contrary, $\mu^{2}$ contains no dependence on $N_{c}$: in
this way the quark-antiquark meson masses scales --as desired-- as $N_{c}^{0}$.

The critical temperature $T_{c}$ for the chiral phase restoration is
calculated by using the so-called CJT formalism \cite{cjt} , which is a
self-consistent resummation scheme for field theoretical calculations at
nonzero temperature. In the Hartree and in the double-bubble approximation
$T_{c}$ is given by the expression
\begin{equation}
T_{c}(N_{c})=f_{\pi}\sqrt{2\frac{N_{c}}{3}}\propto\sqrt{N_{c}}\text{ ,}%
\end{equation}
where $f_{\pi}=92.4~$MeV is the pion decay constant. The scaling of $T_{c}$ is
thus in disagreement with the NJL model \cite{njl} where $T_{c}$ $\propto
N_{c}^{0}$ and with basic expectations \cite{mclerran}. This result is due to
the fact that for $N_{c}\gg3$ a gas of free mesons is realized and thus no
transition takes place. In fact, the mechanism responsible for the restoration
of chiral symmetry in hadronic models is given by mesonic loops, whose effect
vanishes for $N_{c}\gg3$. On the contrary, in the NJL model the restoration of
chiral symmetry is generated by the quark loops, which do not vanish in the
large-$N_{c}$ limit.

The inconsistency between the NJL model and the $\sigma$-model can be easily
solved by replacing
\begin{equation}
\mu^{2}\rightarrow\mu(T)^{2}=\mu^{2}\left(  1-\frac{T^{2}}{T_{0}^{2}}\right)
\end{equation}
(i.e., making it $T$-dependent) where the parameter $T_{0}\simeq\Lambda
_{QCD}\propto N_{c}^{0}$ introduces a new temperature scale. This is in line
with the fact that the $\sigma$-model can be obtained by hadronization of the
NJL model. In this scheme the parameters of the $\sigma$-model turn out to be
temperature-dependent. Note also that the here considered $T^{2}$-behavior
--although naive at the first sight-- has been also obtained in Ref.
\cite{leut}. In this way the critical temperature is modified to
\begin{equation}
T_{c}(N_{c})=T_{0}\left(  1+\frac{1}{2}\frac{T_{0}^{2}}{f_{\pi}^{2}}\frac
{3}{N_{c}}\right)  ^{-1/2}\propto N_{c}^{0}\text{ ,}%
\end{equation}
which is now large-$N_{c}$ independent, just as in the NJL case. For $N_{c}%
=3$, using $T_{0}=\Lambda_{QCD}\simeq225~\text{MeV}$, the critical temperature
$T_{c}$ is lowered to $T_{c}\simeq113~\text{MeV}$. Interestingly, in the
framework of $\sigma$-models with (axial-)vector mesons \cite{denis}, one has
to make the replacement $f_{\pi}\rightarrow Zf_{\pi}$ with $Z=1.67\pm0.2$. In
this way the critical temperature reads $T_{c}\simeq157~$MeV, which is
remarkably close to the lattice results \cite{latt}.

Beyond the phenomenologically motivated modification presented here, one can
go further and couple the present $\sigma$-model (and generalizations thereof)
to the Polyakov loop \cite{dum}. Also in this case \cite{chirallargen} the
critical temperature turns out to be, as desired, independent on $N_{c}$. The
reason for this behavior can be traced back to the fact that the transition of
the Polyakov loop (which describes the confinement-deconfinemet phase
transition) triggers also the restoration of chiral symmetry.

\section{Conclusions}

In this work we have investigated the properties of nuclear matter and chiral
phase transition in the large-$N_{c}$ limit.

We have found that present knowledge on the spectroscopy of scalar mesons
indicates that nuclear matter does not bind for $N_{c}>3.$ Namely, the
nucleon-nucleon attraction in the scalar-isoscalar channel turns out not to be
strong enough to bind nuclei when $\ N_{c}$ is increased. Therefore, nuclear
matter seems to be a peculiar property of our $N_{c}=3$ world.

For what concerns the chiral phase transition at nonzero temperature and zero
density, we have found that care is needed when using effective hadronic
models of the $\sigma$-type. The critical temperature $T_{c}$ does not scale
as expected in the large-$N_{c}$ limit. It is however possible to introduce
simple modifications of chiral hadronic models in such a way that the expected
result $T_{c}\propto N_{c}^{0}$ is recovered.

\section*{Acknowledgments}

The author thanks L. Bonanno, A. Heinz, and D. H. Rischke for cooperation
\cite{lucafra,chirallargen} and S. Lottini, G. Torrieri, G. Pagliara and
H. Warringa for discussions.


\begin{thebibliography}{99}                                                                                               %




\bibitem {witten}E.~Witten,
Nucl.\ Phys.\ B \textbf{160} (1979) 57.
G.~'t Hooft,
Nucl.\ Phys.\ B \textbf{72} (1974) 461.


\bibitem {mclerran}L.~McLerran and R.~D.~Pisarski,
Nucl.\ Phys.\ A \textbf{796} (2007) 83.
G.~Torrieri and I.~Mishustin,
Phys.\ Rev.\ C \textbf{82} (2010) 055202.
S.~Lottini and G.~Torrieri,
arXiv:1103.4824 [nucl-th].
T.~Kojo, Y.~Hidaka, L.~McLerran and R.~D.~Pisarski,
Nucl.\ Phys.\ A \textbf{843} (2010) 37.
C.~Sasaki and I.~Mishustin,
Phys.\ Rev.\ C \textbf{82} (2010) 035204.


\bibitem {lucafra}L.~Bonanno and F.~Giacosa,
arXiv:1102.3367 [hep-ph].


\bibitem {amslerrev}C.~Amsler and N.~A.~Tornqvist,
Phys.\ Rept.\ \textbf{389}, 61 (2004).
E.~Klempt and A.~Zaitsev,
Phys.\ Rept.\ \textbf{454} (2007) 1.
F.~Giacosa,
Phys.\ Rev.\ D \textbf{80} (2009) 074028.


\bibitem {njlorig}Y.~Nambu, G.~Jona-Lasinio,
Phys.\ Rev.\ \textbf{122}, 345-358 (1961);
Y.~Nambu, G.~Jona-Lasinio,
Phys.\ Rev.\ \textbf{124}, 246-254 (1961).


\bibitem {njl}S.~P.~Klevansky,
Rev.\ Mod.\ Phys.\ \textbf{64} (1992) 649.


\bibitem {geffen}M. Gell-Mann, M. Levy,
Nuovo\ Cimento,\ \textbf{16}, 705 (1960).
S.~Gasiorowicz, D.~A.~Geffen,
Rev.\ Mod.\ Phys.\ \textbf{41}, 531-573 (1969).


\bibitem {denis}D.~Parganlija, F.~Giacosa, D.~H.~Rischke,
Phys.\ Rev.\ \textbf{D82}, 054024 (2010).
S.~Gallas, F.~Giacosa, D.~H.~Rischke,
Phys.\ Rev.\ \textbf{D82 } (2010) 014004.
S.~Janowski, D.~Parganlija, F.~Giacosa and D.~H.~Rischke,
arXiv:1103.3238 [hep-ph].




\bibitem {walecka}J.~D.~Walecka,
Annals Phys.\ \textbf{83}, 491 (1974).
B.~D.~Serot and J.~D.~Walecka,
Adv.\ Nucl.\ Phys.\ \textbf{16}, 1 (1986).
B.~D.~Serot and J.~D.~Walecka,
Int.\ J.\ Mod.\ Phys.\ E \textbf{6}, 515 (1997).


\bibitem {jaffeorig}R.~L.~Jaffe,
Phys.\ Rev.\ D \textbf{15} (1977) 267.
L.~Maiani, F.~Piccinini, A.~D.~Polosa and V.~Riquer,
Phys.\ Rev.\ Lett.\ \textbf{93} (2004) 212002 [arXiv:hep-ph/0407017].
F.~Giacosa,
Phys.\ Rev.\ D \textbf{74} (2006) 014028.
F.~Giacosa,
Phys.\ Rev.\ D \textbf{75} (2007) 054007.
A.~H.~Fariborz, R.~Jora and J.~Schechter,
Phys.\ Rev.\ D \textbf{72} (2005) 034001.




\bibitem {heinz}A.~Heinz, S.~Struber, F.~Giacosa and D.~H.~Rischke,
Phys.\ Rev.\ D \textbf{79} (2009) 037502.
S.~Gallas, F.~Giacosa and G.~Pagliara,
arXiv:1105.5003 [hep-ph].




\bibitem {weisenucl}N.~Kaiser, R.~Brockmann and W.~Weise,
Nucl.\ Phys.\ A \textbf{625} (1997) 758.
A.~Calle Cordon and E.~Ruiz Arriola,
Phys.\ Rev.\ C \textbf{81} (2010) 044002.


\bibitem {varieglue}
C.~Amsler and F.~E.~Close,
Phys.\ Rev.\ D \textbf{53} (1996) 295.
W.~J.~Lee and D.~Weingarten,
Phys.\ Rev.\ D \textbf{61}, 014015 (2000).
F.~E.~Close and A.~Kirk,
Eur.\ Phys.\ J.\ C \textbf{21}, 531 (2001).
F.~Giacosa, T.~Gutsche, V.~E.~Lyubovitskij and A.~Faessler,
Phys.\ Rev.\ D \textbf{72}, 094006 (2005);
Phys.\ Lett.\ B \textbf{622}, 277 (2005).
H.~Y.~Cheng, C.~K.~Chua and K.~F.~Liu,
Phys.\ Rev.\ D \textbf{74} (2006) 094005.
L.~Bonanno and A.~Drago,
Phys.\ Rev.\ C \textbf{79}, 045801 (2009).
V.~Mathieu, N.~Kochelev and V.~Vento,
Int.\ J.\ Mod.\ Phys.\ E \textbf{18} (2009) 1.


\bibitem {lattglue}Y.~Chen \textit{et al.},
Phys.\ Rev.\ D \textbf{73} (2006) 014516.


\bibitem {rischkerev}D.~H.~Rischke,
Prog.\ Part.\ Nucl.\ Phys.\ \textbf{52} (2004) 197.
E.~Megias, E.~Ruiz Arriola, L.~L.~Salcedo,
Phys.\ Rev.\ \textbf{D74}, 065005 (2006).


\bibitem {cjt}
J.~M.~Cornwall, R.~Jackiw, E.~Tomboulis,
Phys.\ Rev.\ \textbf{D10}, 2428-2445 (1974).


\bibitem {leut}J.~Gasser, H.~Leutwyler,
Phys.\ Lett.\ \textbf{B184}, 83 (1987).


\bibitem {latt}M.~Cheng, N.~H.~Christ, S.~Datta, J.~van der Heide, C.~Jung,
F.~Karsch, O.~Kaczmarek, E.~Laermann \textit{et al.},
Phys.\ Rev.\ \textbf{D74 } (2006) 054507.


\bibitem {dum}A.~Dumitru, R.~D.~Pisarski,
Phys.\ Lett.\ \textbf{B504 } (2001) 282-290.




\bibitem {chirallargen}A. Heinz, F. Giacosa, and D. H. Rischke, in preparation.
\end{thebibliography}
\end{document}